**[Title page]**

**[Article Full Title]**

# A Novel Automatic Real-time Motion Tracking Method in MRI-guided Radiotherapy Using Enhanced Tracking-Learning-Detection Framework with Automatic Segmentation

**[Author Names]**

Shengqi Chen[1], B.Eng; (Email: shengqichen@bupt.edu.cn )

Zilin Wang[1], B.Eng; (Email: roff972165375@163.com )

Jianrong Dai[2], Ph.D; (Email: dai_jianrong@cicams.ac.cn )

Shirui Qin[2], MSc; (Email: shiruiok@126.com )

Ying Cao[2], B.S; (Email: kugua1225@126.com )

Ruiao Zhao[2], B.S; (Email: 13115331658@163.com )

Jiayun Chen[2]*, Ph.D; (Corresponding author, Email: grace_chenjy@163.com)

Guohua Wu[1]*, Ph.D; (Corresponding author, Email: wuguohua@bupt.edu.cn)

Yuan Tang[2]*, MD (Corresponding author, Email: tangyuan82@126.com )

*These authors contributed equally to this work. Jiayun Chen, Guohua Wu, and Yuan Tang contributed equally as corresponding authors

**[Author Institutions]**

[1] School of Electronic Engineering, Beijing University of Posts and Telecommunications, 10 Xitucheng Road, Haidian District, Beijing, 100876, China

[2] Department of Radiation Oncology, National Cancer Center/National Clinical Research Center for Cancer/Cancer Hospital, Chinese Academy of Medical Sciences and Peking Union Medical College, 17 Panjiayuan South Road, Chaoyang District, Beijing, 100021, China

**[Corresponding Author Name & Email Address]**

Jiayun Chen[2]*, Ph.D; (Corresponding author, Email: grace_chenjy@163.com)

Guohua Wu[1]*, Ph.D; (Corresponding author, Email: wuguohua@bupt.edu.cn)

Yuan Tang[2]*, MD (Corresponding author, Email: tangyuan82@126.com )

**[Abstract]**

**Background and purpose:** Accurate motion tracking in magnetic resonance imaging-guided radiotherapy (MRIgRT) is essential for effective treatment delivery. This study aimed to enhance motion tracking precision in MRIgRT through an automatic real-time markerless tracking method using an enhanced Tracking-Learning-Detection (ETLD) framework combined with automatic segmentation, eliminating the need for pre-training.

**Methods:** We developed a novel motion tracking and segmentation method by integrating the ETLD framework with an improved Chan-Vese (ICV) model, termed ETLD+ICV. The ETLD framework was upgraded for real-time MRIgRT, including search process optimization, an enhanced median-flow tracker, and dynamic detection region adjustments. It receives 3.5D MRI data as input and outputs the location prediction of the target volume for each frame. Based on this, ICV was used for precise target volume coverage, refining the segmented region frame by frame using tracking results, with optimized key parameters. The method requires no pre-training and was tested on 3.5D MRI scans from 10 patients with liver metastases. No-reference image quality assessment and manual verification were used to filter out low-quality image data. Comprehensive statistical analyses were performed to assess the statistical significance of differences between experimental outcomes ($p < 0.05$).

**Results:** Evaluation across 106,000 frames from 77 treatment fractions demonstrated sub-millimeter tracking errors of less than 0.8 mm, with over 99% precision and 98% recall for all patients in the Beam Eye View (BEV)/Beam Path View (BPV) orientation. The ETLD+ICV method achieved a Dice global score of more than 82% for all patients, demonstrating the method's extensibility and precise target volume coverage.

**Conclusion:** This study successfully developed an automatic real-time markerless motion tracking method for MRIgRT that outperforms current methods. The novel method not only delivers exceptional precision in tracking and segmentation but also shows enhanced adaptability to clinical demands, making it an indispensable asset in improving the efficacy of radiotherapy treatments.

## 1. Introduction

Cancer has emerged as a significant health challenge among non-communicable diseases, impacting a substantial portion of the global population. Radiotherapy, recognized as one of the most effective clinical treatments for cancer, offers several advantages, including efficient tumor eradication, protecting healthy organs, and non-invasiveness. Approximately 70% of cancer patients receive radiotherapy at some stage of their treatment(1), with the primary objective of delivering high radiation doses to the tumor while minimizing exposure to surrounding healthy tissues(2, 3). However, physiological processes such as respiration and gastrointestinal activity can induce tumor position shifts from 2 to 14 mm, potentially compromising the accuracy of radiation delivery and increasing the dose to healthy tissues, thereby affecting treatment efficacy(2, 4, 5). A wealth of evidence from previous studies highlights this point. For instance, in the abdominal regions, one study demonstrated that even minor intra-fraction motion (mean error < 8mm) can result in a significant reduction of up to 24% in the radiation dose to the liver(6). Another study reported that in the absence of motion adaptation, for a planning target volume (PTV) with a cranio-caudal margin of 7 mm, liver tumors with an average geometric error of less than 10 mm experienced a radiation dose reduction of over 29%(7). Similarly, research on the pancreatic tumors revealed that mean motion amplitudes of 9.6 mm in the cranio-caudal direction and 6.9 mm in the anterior-posterior direction can lead to a 28% reduction in the prescribed dose to the clinical target volume (CTV)(8). Accurate compensation for these shifts is crucial to ensure that the tumor receives the prescribed dose while minimizing the impact on healthy tissues, making the management of intra-fraction motion a critical focus for improving patient outcomes(5, 9).

Image-guided radiotherapy (IGRT) is a cornerstone for precision radiotherapy, with cone-beam CT (CBCT) being widely used for online tumor positioning verification(10). However, CBCT has several limitations, including prolonged scanning times and suboptimal tissue contrast(11). Tumor surrogates, such as the diaphragm and chest wall, have been employed for tracking(12), but the correlation between tumor motion and these surrogates can vary significantly among individuals(13). Invasive methods, such as implantation of gold fiducials within tumors, can introduce discomfort and uncertainties(3, 14).

Magnetic resonance imaging-guided radiotherapy (MRIgRT) has emerged as a promising alternative, offering superior soft tissue contrast and real-time imaging capabilities. These features facilitate accurate tumor targeting and help reduce radiation toxicity to surrounding tissues(10, 15). Despite its relatively recent clinical adoption, MRIgRT has demonstrated substantial benefits for cancer patients(16, 17). Accelerated imaging techniques have made MRIgRT viable for online treatment plan adaptation. The MRI-Linac (MRI-guided linear accelerator) system provides comprehensive information throughout the treatment process, supporting both manual and automatic decision-making(15, 18, 19). It is anticipated that MRIgRT will significantly contribute to the advancement of IGRT in the future.

During MRIgRT, physicians can dynamically adjust the clinical target volume (CTV) and plan target volume (PTV) online(15). Currently, the standard practice for creating a PTV is to expand the CTV with a spatial margin to account for setup uncertainties and organ deformations. Therefore, accurate image acquisition and tumor motion monitoring are crucial for ensuring the effectiveness and safety of the treatment(20). However, imaging artifacts and deformations can impair physicians' recognition and interpretation of the CTV and PTV(21). Moreover, current systems often rely on manual contouring, which is inefficient and prone to inconsistencies(22). Studies have investigated automatic, markerless motion tracking based on MRI to address these challenges, offering the advantage of being non-invasive and beneficial to patients. In early studies, Yun J et al. developed a lung tumor tracking algorithm based on the pulse coupled neural network (PCNN), which achieved millimeter-level tracking errors. However, its clinical applicability was limited by requiring manual pretreatment procedures and patient-specific parameter optimization[(23, 24)]. Bourque A E et al. proposed a motion prediction scheme combining the dynamic autoregressive model with a particle filter algorithm, achieving similar millimeter-level accuracy in lung tumors. This method, however, required manual delineation of regions of interest (ROI) for reference template generation, potentially introducing subjective errors(25, 26). Seregni M et al. developed a hybrid framework integrating optical flow methods with template matching for real-time liver motion tracking(27). Although balancing efficiency and precision, its static template design showed limited adaptability to respiratory motion dynamics and baseline drift. Despite their innovative contributions, these early approaches exhibited inherent limitations that hindered their seamless clinical translation, necessitating further refinement to enhance their practicality and reliability in clinical settings.

Deep learning (DL) models, possessing powerful feature extraction capabilities, have been extensively utilized for motion tracking in MRIgRT. Terpstra M L et al. introduced a multi-resolution convolutional neural network (CNN) named TEMPEST, designed to estimate deformation vector fields (DVF). This model achieved tracking errors below 2 mm and demonstrated competitive performance on external test sets(28). Shao H C et al. developed an unsupervised motion estimation model called KS-RegNet, based on the U-Net architecture. This obtained Dice similarity coefficients (DSC) of 0.894 and 0.766 for cardiac and abdominal datasets, respectively, while showing robustness across different under-sampling factors(29). Hunt B et al. implemented cine-MRI sequence registration using VoxelMorph for real-time target motion tracking, demonstrating significant superiority over conventional methods(30). Peng J employed U-Net and its variants to achieve superior performance in lung tumor tracking with a maximum mean DSC of 0.86, outperforming generic onboard algorithms (DSC=0.68)(31). However, these studies also emphasize the critical importance of data in motion tracking. When training data is limited, DL methods may underperform compared to traditional approaches, highlighting the necessity of large-scale training datasets(30). Peng J et al. further revealed that the influence of data on tracking performance outweighs differences in model architecture(31). For DL models, acquiring high-quality annotated training data remains a significant challenge and constitutes a practical limitation(28). Meanwhile, inconsistencies in manually labeled ground truth can also impose constraints on the deep learning training process(3). These limitations indicate that tracking methods must reduce their reliance on imaging data while maintaining accuracy, a challenge that stems from the inherent scarcity of medical imaging data.

To address the challenges mentioned above, this study introduces the Tracking-Learning-Detection (TLD) framework(32), a precise object tracking method that requires no pre-training. This framework has been successfully applied in various computer vision tasks, such as moving vehicle tracking(33) and pedestrian tracking(34). DHONT J et al. have demonstrated its effectiveness in MRIgRT applications, exhibiting adaptability to the target volume's disappearance and deformation induced by irregular respiratory motion(35). However, it should be noted that they also reported non-negligible tracking failures in 0.35T MRI data, underscoring the need for targeted enhancements to improve the algorithm's applicability in clinical radiotherapy.

This study aims to develop and validate an enhanced real-time motion tracking approach for MRIgRT by enhancing the TLD framework and integrating it with an improved Chan-Vese (ICV) model for precise target volume coverage. Specifically, the primary contributions of this study can be summarized as follows: (1) We proposed the enhanced Tracking-Learning-Detection (ETLD) algorithm, which incorporates three critical enhancements: search process optimization, median-flow tracker enhancement, and dynamic detection region adjustment, which enable high-precision real-time motion tracking in MRIgRT. (2) By optimizing key parameters of the Chan-Vese model, we developed an ICV model. The integration of ETLD with ICV (ETLD+ICV) enabled precise target volume coverage. (3)Validation using 3.5D MRI datasets from a cohort of 10 patients with liver metastases demonstrated that ETLD achieves sub-millimeter tracking errors, while ETLD+ICV maintains accurate target volume segmentation. The manuscript is organized as follows: Section 2 details the materials and methods, including data acquisition protocols and the specifics of the proposed method. Section 3 presents the experimental results. Section 4 discusses our work and highlights the key findings. Finally, Section 5 summarizes the core innovations of this study.

## 2. Materials and Methods

### 2.1. High Spatial and Temporal Resolution MRI Method with BEV/BPV Fusion Information

In our previous work, we developed a novel high-resolution fast MRI method, designated as 3.5D MRI(36), which integrates Beam Eye View (BEV) and Beam Path View (BPV) information. This method optimizes MRI centroid alignment with the radiation beam's center. It generates three orthogonal imaging planes: two BPV planes aligned with the radiation rays and one BEV plane perpendicular to them. These planes are dynamically adapted to the tumor's centroid and the orientation of the radiation field. Importantly, this method enables real-time localization of both the target volume and organs at risk, eliminating the need for post-processing adjustments.

## 2.2. Data Acquisition and Preprocessing

From August 2021 to February 2024, patients with colorectal cancer liver metastasis who received treatment using the MR-Linac Unity (Elekta AB, Stockholm, Sweden) at our department provided informed consent for online MRI-guided radiotherapy (MRIgRT). The treating radiation oncologist ensured that there were no contraindications for radiotherapy.

For 4D-CT scanning (Siemens Healthcare, Erlangen, Germany), patients were positioned supine with arms raised and supported on a headrest bracket. Target volume delineation was performed based on the scanned images, and the treatment plan was devised using the Monaco Version 5.4 system (Elekta AB, Stockholm, Sweden) for intensity-modulated radiotherapy (IMRT), employing 6-9 beams. During treatment, patients followed a standard protocol using an abdominal compression belt to limit abdominal movement, and 2D cine-MRI was utilized for continuous tumor position monitoring to ensure dose delivery accuracy. After each treatment, the abdominal compression belt was reapplied to restrict abdominal movement, and 3.5D MRI data (group RAM) was acquired. Subsequently, patients were scanned in a free-breathing state to obtain 3.5D MRI data (group FB). Detailed acquisition protocols and acquisition parameters are available in our previous research(36).

The study included a total of 10 patients, with 3.5D MRI collected from multiple treatment fractions, captured frame by frame across one BEV plane (transverse) and two BPV planes (coronal and sagittal) intersecting at the treatment isocenter with 50 frames per plane at a frequency of 4.347 Hz. By selecting 2D cine-MRI images from the coronal and sagittal planes, we compiled an MRI dataset comprising 112,800 frames from 77 treatment fractions, totaling 2,256 subsets. The resolution of each frame ranges from 320×320 to 560×560 pixels, with pixel spacing varying between 0.571×0.571 and 0.938×0.938 mm. Variability was observed in the angles between the longitudinal cine-MRI imaging and coronal planes, as well as in imaging intervals across different patients. Table 1 lists the detailed information of our datasets, including patient-specific treatment fractions, radiation field angles, and the corresponding number of data groups.

**Table 1. Details of datasets.**

| Patient | No. of fractions | Beam angles | No. of groups |
|---|---|---|---|
| 1 | 4 | 140,160,180,200,220,240,260,280,300 | 144 |
| 2 | 10 | 40,70,130,220,250,280,310,340 | 320 |
| 3 | 5 | 120,150,180,210,260,290,320,350 | 160 |
| 4 | 8 | 210,230,250,300,340 | 160 |
| 5 | 10 | 0,25,50,125,150,175,335 | 280 |
| 6 | 5 | 160,185,210,235,260,285,310,335 | 160 |
| 7 | 13 | 5,20,35,50,65,240,255,270 | 416 |
| 8 | 8 | 24,160,270,290,310,330,350 | 224 |
| 9 | 9 | 180,210,235,260,285,310,330 | 252 |
| 10 | 5 | 0,30,55,80,130,190,340 | 140 |
| Sum | 77 | | 2256 |

To enhance image quality and ensure compliance with TLD tracking assumptions, all cine-MRI images underwent global grayscale normalization to address brightness inconsistencies. Gamma correction was introduced to enhance contrast, while a Gaussian filter was applied to smooth out noise. A no-reference image quality assessment (NRIQA) approach was implemented to evaluate preprocessing efficacy and determine optimal parameters quantitatively. Specifically, we utilized the BRISQUE model, an NRIQA method that evaluates deviations from natural imaging scene statistics in image features to compute quality scores(37). This methodology inherently eliminates the dependency on high-quality reference data, offering a critical solution for medical imaging scenarios where high-quality reference images are usually unavailable. We incorporated median absolute deviation (MAD) weighting to strengthen this quality assessment framework for dynamic image quality evaluation(38). Furthermore, to maintain dataset validity, cine-MRI sequences with mean quality scores below the threshold were excluded as low-quality data, where the threshold was empirically set to the 5th percentile of the scores. A radiation oncologist conducted secondary verification of the entire process to validate the preprocessing protocol's effectiveness and prevent erroneous data exclusion outcomes.

### 2.3. Enhanced Tracking-Learning-Detection (ETLD)

The TLD framework is a machine learning algorithm designed for long-term tracking of a single object within video streams(32). It comprises three core components: (1) a tracker performing inter-frame motion estimation under the assumption of limited object displacement and continuous visibility, (2) a detector conducting global searches for target regions based on historical detections and object model, and (3) a learning component that evaluates detection errors to refine the object model through self-generated training samples iteratively. The framework receives image sequences as input and directly outputs predicted target bounding boxes, eliminating additional post-processing operations. Crucially, tracking integration with detection enables TLD to handle target disappearance and reappearance while adapting to appearance changes through its online-updated object model and learning component, achieving high-performance object tracking without requiring pre-training. These characteristics establish TLD as an ideal solution for addressing target volume motion tracking in MRIgRT.

However, TLD is typically deployed in natural imaging scenarios with higher spatial resolution and signal-to-noise (SNR) ratios than cine-MRI, where high-quality video streams provide richer target features. Therefore, direct migration this framework to MRI data may encounter challenges such as degraded feature discriminability, potentially leading to tracking failure(35). Additionally, quantitative analysis demonstrated that over 97.9% of tracked targets in this study (as defined in Section 2.5) occupied less than 1% of the global image, exhibiting a substantial disparity compared to targets in conventional tracking scenarios(39). This emphasizes the necessity for specialized optimizations addressing small-target characteristics.

To address these challenges, we have enhanced the original TLD (OTLD) framework, termed the enhanced TLD (ETLD), as illustrated in Figure 1. The ETLD accepts preprocessed 2D cine-MRI sequences as input and processes them through tracking, detection, learning, and integration modules to generate unified location predictions. Three critical enhancements specifically designed for real-time MRIgRT are highlighted in Figure 1 (marked with red rounded rectangles) and are detailed as follows:

1) Search Process Optimization: We optimized the search process parameters by reducing the scanning window size from 24×24 pixels to 12×12 pixels and the image patch size from 15×15 pixels to 7×7 pixels, resulting in a reduction to approximately one quarter of the original size to better accommodate small target characteristics. Additionally, given that target size typically demonstrates minimal variation, the sliding window scaling factor was adjusted from 1.2 to 1.1 to capture subtle scale changes precisely.

2) Median-Flow Tracker Enhancement: The median-flow tracker in OTLD operates under the spatial consistency assumption within the search neighborhood. Based on a previous study demonstrating a maximum 95th percentile liver tumor displacement of 10 mm(36), we expanded the tracker's search neighborhood from 4×4 to 30×30 pixels and increased the iteration count to 30 to mitigate tracking errors.

3) Dynamic Detection Region Adjustment: Motivated by the same consideration, we restricted the detector's search region from a global search to an online-updated region. Specifically, we set the search region as a 30×30 pixels region centered on the average coordinates of the target's center from the previous three frames, thereby avoiding potential false results.

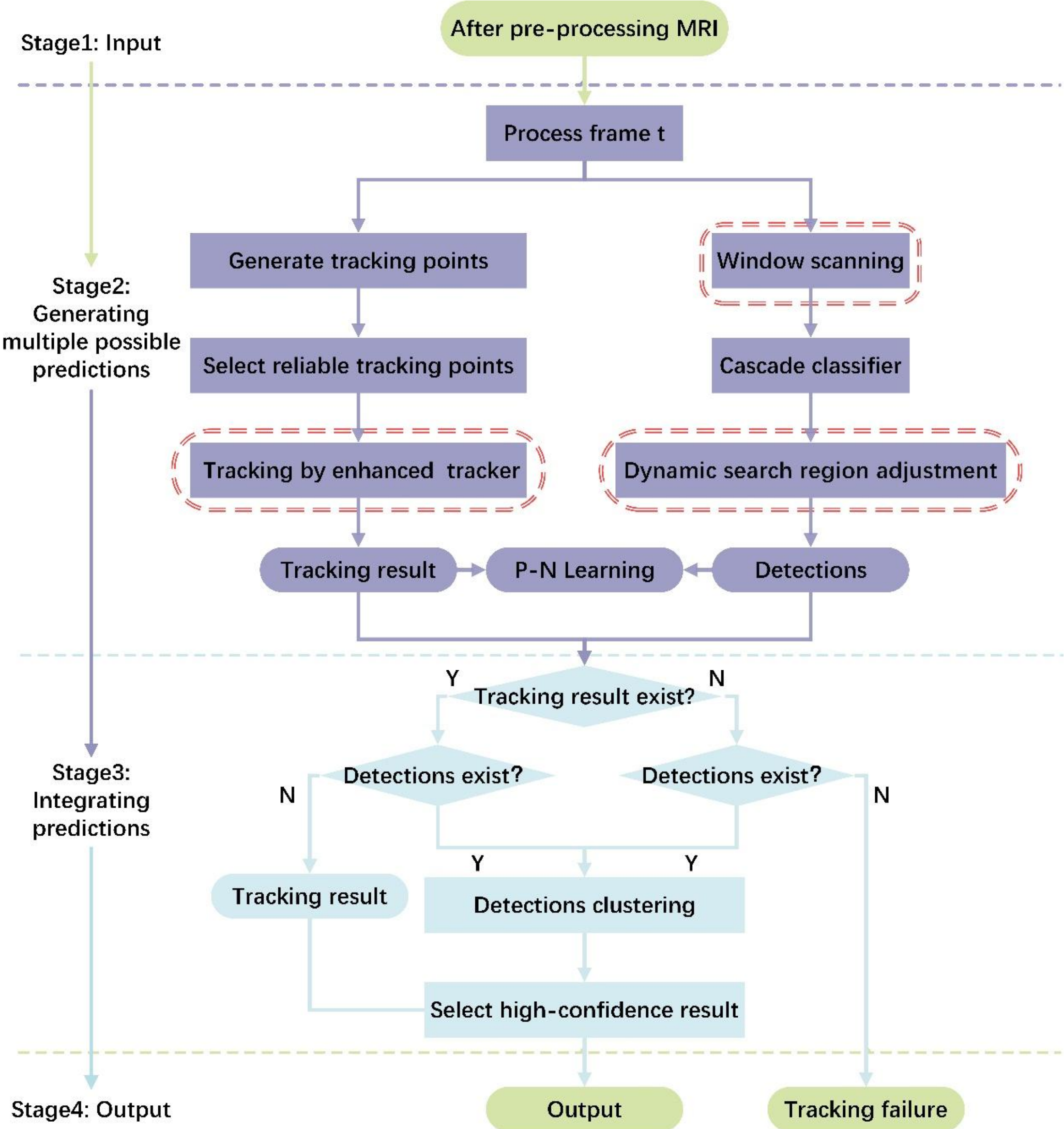

**Figure 1. The main workflow of the ETLD method. The red rounded rectangles denote**

**the main improvements based on the OTLD.**

### 2.4. Automatic Segmentation Process

Based on the location predictions provided by the ETLD method, we integrated a segmentation process to ensure accurate radiotherapy beam coverage of the ROI. This was achieved by combining the ETLD method with the Chan-Vese model(40). The Chan-Vese model delineates the target by iteratively minimizing an energy function with regularization terms and has been used in medical image segmentation tasks(41).

In a typical implementation, the Chan-Vese model takes a global image sequence and a fixed mask to achieve contour segmentation. For our specific implementation, the Chan-Vese model receives the cine-MRI sequence together with an initial bounding box as inputs. In each frame, the segmentation model leverages both the initial bounding box and the real-time location prediction from the ETLD method to refine the contour through energy minimization iteratively. This synergistic approach allows precise monitoring of the target volume's morphological and positional variations throughout treatment. Furthermore, we optimized the model parameters to ensure segmentation accuracy while meeting real-time performance requirements. For ease of reference, we have designated our integrated approach, which combines the ETLD framework with the ICV model, as the ETLD+ICV method. This enhances the precision of radiotherapy beam targeting and ensures clinical reliability.

### 2.5. Evaluation Procedure

The proposed method, implemented in C++ and MATLAB (2021b), was tested on a personal computer (PC) with an Intel(R) Core(TM) i9-13980HX processor and 32GB RAM. Notably, our method's efficiency eliminates the need for graphic processing unit (GPU) acceleration, making it highly accessible. All experiments were conducted in a Windows 11 environment with Visual Studio 2019 and OpenCV 3.3.1.

Since neither the TLD algorithm nor the Chan-Vese model requires training, data from all patients were utilized to evaluate the proposed method's performance, encompassing 10 patients and 112,800 frames. Given the subtlety of tumor delineation in 3.5D MRI, we opted to track blood vessels as a proxy for tumor motion. This strategy is supported by a robust body of evidence demonstrating a strong positive correlation between liver tumors and blood vessel motion(42, 43), rendering blood vessels a reliable surrogate for our study.

The ground truth was established through manually delineated binary masks of tumor regions by a resident physician using ITK-SNAP (version 4.0.1) over 9 months. For rigorous evaluation, all MRI data and delineation results underwent dual validation based on the NRIQA model (Section 2.2) and manual verification by a senior radiation oncologist. Low-quality images were excluded, and incorrect delineations were corrected, resulting in a refined dataset of 2,120 subsets and 106,000 frames.

### 2.6. Evaluation Metrics

The effectiveness of our proposed method is rigorously evaluated using a comprehensive set of metrics, detailed as follows:

#### 2.6.1. Tracking Performance Evaluation

Accurate tracking of the tumor center's trajectory by the radiation beam is crucial in clinical settings. To reflect this priority, the real-time trajectory of the radiotherapy beam was simulated using the output of the tracking algorithm. The mean absolute error (MAE) is the primary metric for this assessment, as shown in Equation (1):

$$MAE = \frac{1}{N}\sum_{i=1}^{N}\left|\vec{P}_i - \vec{G}_i\right| \quad (1)$$

Here, $\vec{P}_i$ and $\vec{G}_i$ represent the predicted and actual displacement vectors at frame $i$, respectively, with $|\cdot|$ signifying the absolute value, and $N$ is the total number of frames. The MAE quantifies the tracking algorithm's adherence to the tumor's motion.

We computed the root mean square error (RMSE) between predictions and ground truth to further substantiate the tracking efficacy. The correlation coefficient (CC) was also employed to validate tracking efficacy, with higher CC values indicating better performance(44). Precision and recall rates were also calculated across the dataset to evaluate clinical applicability.

### 2.6.2. Segmentation Performance Evaluation

For the segmentation facet, the Dice coefficient, a standard metric in the field, is employed as depicted in Equation (2):

$$Dice = \frac{2\left|mask_i^{prediction} \cap mask_i^{groundtruth}\right|}{\left|mask_i^{prediction}\right| + \left|mask_i^{groundtruth}\right|} \quad (2)$$

Here, $mask_i^{prediction}$ and $mask_i^{groundtruth}$ denote the predicted and actual segmentation masks, respectively, with $\cap$ representing the intersection and $|\cdot|$ quantifies the pixel count. Given the varying target sizes and the Dice coefficient's sensitivity(45), the Dice global metric from the Liver Tumor Segmentation Challenge (LiTS)(46) was applied to assess segmentation efficacy, capturing performance across varying target sizes and contour estimation sensitivity.

### 2.6.3. Computational Efficiency Evaluation

The metric of frames per second (FPS) encapsulates the method's real-time capabilities. Tracking and segmentation FPS were calculated for each patient to present the computational efficiency of the method, ensuring compliance with real-time clinical radiotherapy constraints.

## 2.7. Statistical Analysis

We performed comprehensive statistical analyses on primary metrics to assess the statistical significance of differences between experimental outcomes, with $p < 0.05$ considered statistically significant. Data normality and homogeneity of variance were evaluated using the Shapiro-Wilk test and Levene's test, respectively. The parametric student's t-test was applied when both assumptions were satisfied, while the non-parametric Mann-Whitney U-test was employed otherwise. All statistical analyses were conducted using SPSS version 27.0.1.0 (IBM, Armonk, NY, USA).

# 3. Results

## 3.1. Accuracy of Tracking

Figure 2 illustrates the exemplary tracking outcomes of the ETLD method on the whole treatment course within the RAM group of one selected patient at a gantry angle of 140 degree. The displacement changes are depicted in the graph, with the red line signifying the actual trajectory and the blue line indicating the algorithm's predictions. The juxtaposed images showcase representative frames, highlighting the minimal divergence between predictions and ground truth, thus evidencing a robust correlation.

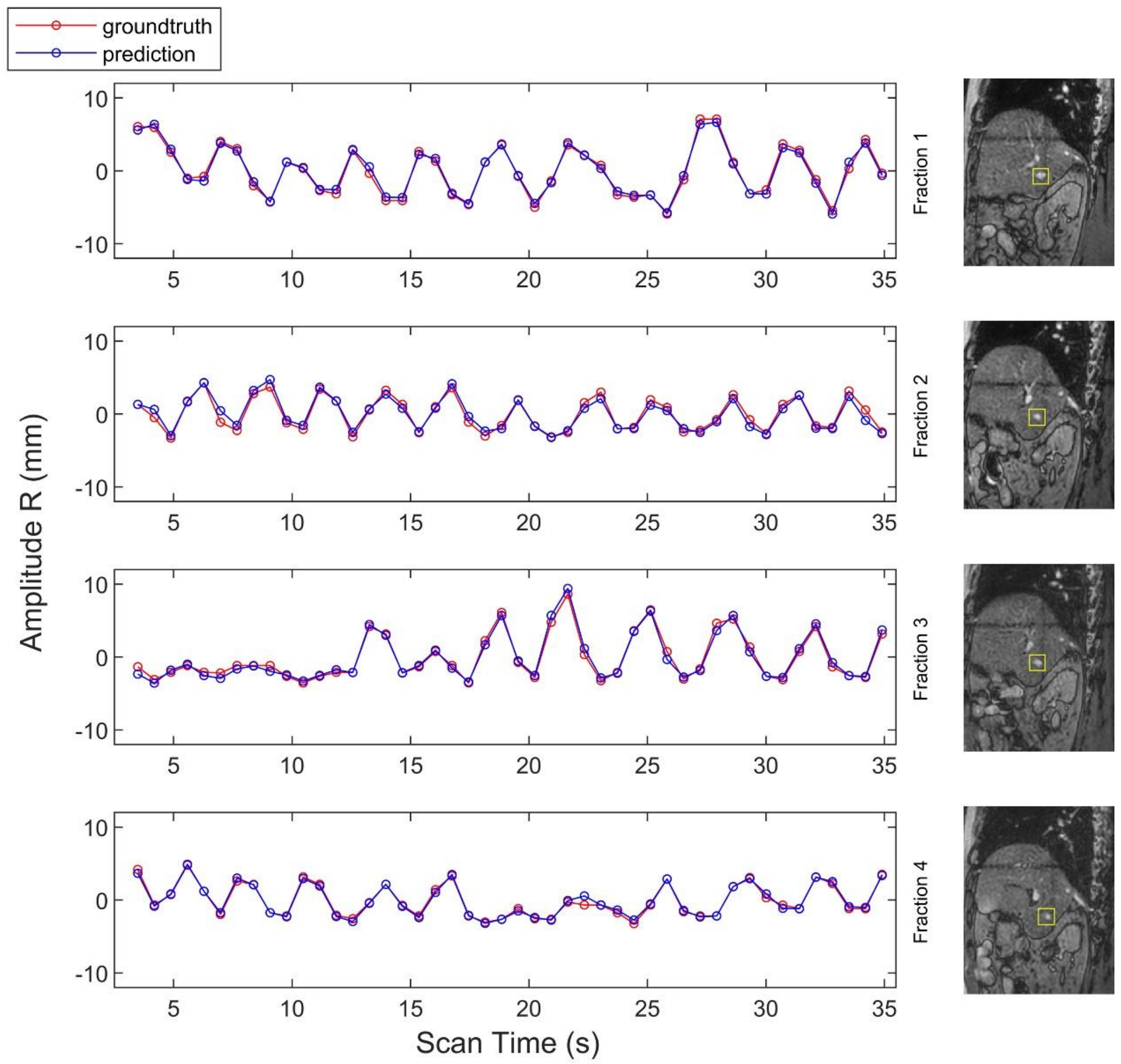

**Figure 2. Tracking results of the ETLD on patient 1 (RAM group) at a 140-degree angle. The graph juxtaposes the actual tumor displacement (red line) with the predicted tracking (blue line), demonstrating the close correlation between the two. Accompanying images showcase the precision of tracking in representative frames, with minimal divergence between the ground truth and predictions, indicative of the algorithm's high fidelity in motion estimation.**

Figure 3 and Table 2 present a comparative analysis of the tracking performance between our proposed ETLD and the OTLD method across the entire dataset. The MAE box plots in Figure 3(a) demonstrate that the ETLD achieves motion tracking errors mostly within 1 mm, outperforming OTLD. As MAE calculates average errors, it reduces the impact of tracking failure frames, leading to partial overlap in box plot distributions. Therefore, precision and recall curves were plotted in Figures 3(b) and (c), demonstrating ETLD's clear superiority. The precision curve indicates evident prediction errors in OTLD for patients 3 to 7, while the recall curve reveals remarkable tracking failures in OTLD, most notably for patient 6.

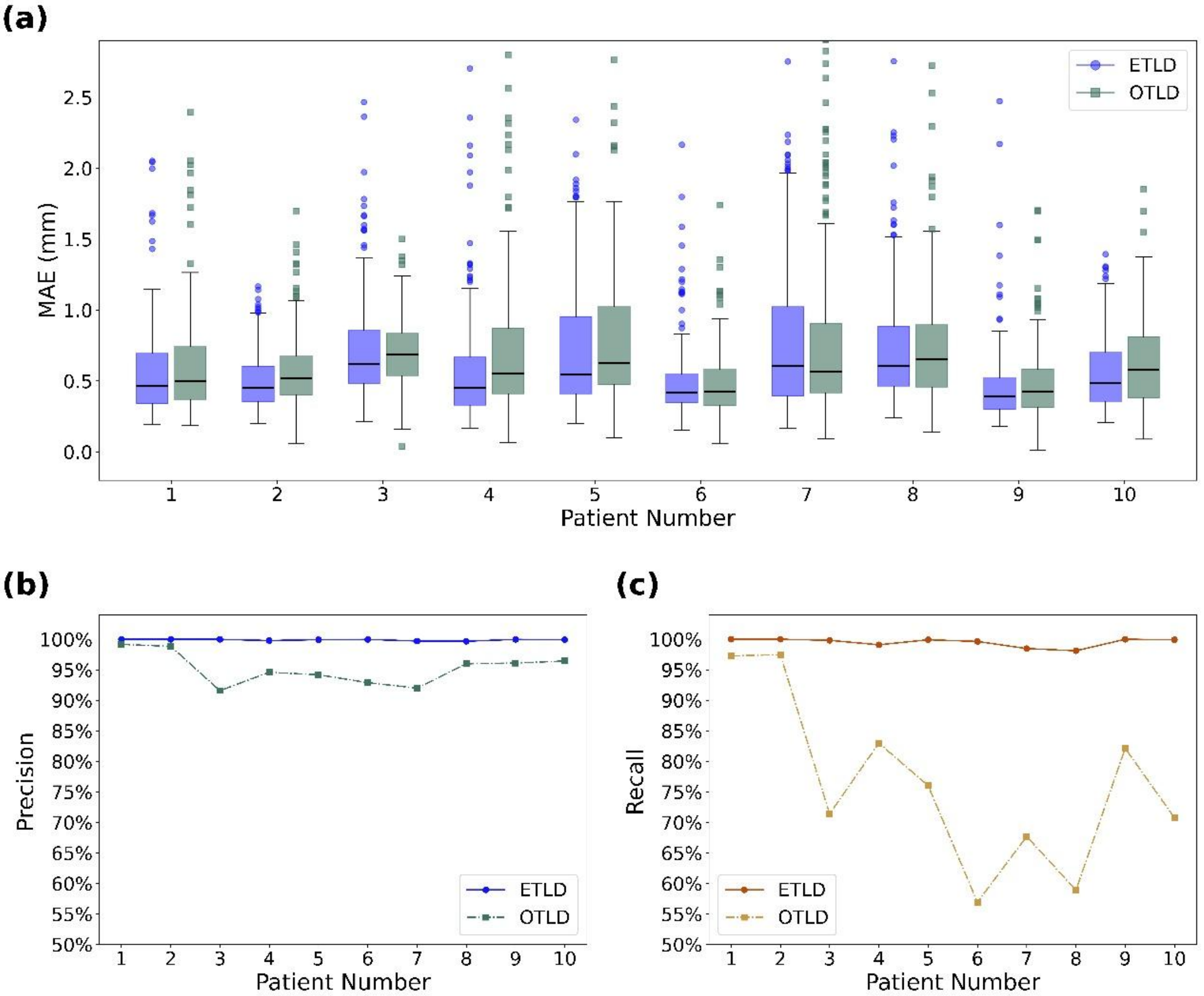

**Figure 3. Comparative tracking performance evaluation of the ETLD and the OTLD methods across 10 patients. (a) The box plot of the MAE in tracking accuracy. (b) The line plot of the precision in tracking. (c) The line plot of the recall in tracking.**

Table 2 lists the quantitative tracking results. The ETLD framework achieved MAE below 0.8 mm, RMSE below 1.05 mm, and CC exceeding 94.3% across all patients, demonstrating superior performance over OTLD in all metrics except for Patient 6. This exception stems from OTLD's inability to track nearly half of the frames in this patient's dataset, thereby excluding high-error frames from error calculations. Notably, ETLD maintained precision above 99% and recall above 98% for all patients. Overall, ETLD achieved MAE lower than 0.65 mm, RMSE below 0.85 mm, CC values surpassing 96%, with both precision and recall exceeding 99%. In contrast, OTLD exhibited precision under 96% and recall below 77%. Further analysis identified 158 complete tracking failures (defined as 2D cine-MRI sequences in which no frames were successfully tracked ) in OTLD across the 2,120 cine-MRI subsets, while no such failures were observed in ETLD.

**Table 2. Comparative analysis of tracking performance metrics between the ETLD and OTLD algorithms across 10 patients. Metrics include mean absolute error (MAE), root mean square error (RMSE), correlation coefficient (CC), precision, and recall.**

| Patient | MAE (mm) | | RMSE (mm) | | CC | | Precision | | Recall | |
|---|---|---|---|---|---|---|---|---|---|---|
| | ETLD | OTLD | ETLD | OTLD | ETLD | OTLD | ETLD | OTLD | ETLD | OTLD |
| 1 | 0.602±0.495 | 0.664±0.521 | 0.813±0.652 | 0.883±0.674 | 94.3% | 93.6% | 100.0% | 99.2% | 100.0% | 97.3% |
| 2 | 0.493±0.188 | 0.565±0.238 | 0.643±0.294 | 0.743±0.371 | 98.5% | 98.0% | 100.0% | 98.8% | 100.0% | 97.5% |
| 3 | 0.747±0.478 | 0.760±0.718 | 0.948±0.586 | 0.978±0.856 | 97.3% | 97.0% | 100.0% | 91.6% | 99.8% | 71.4% |
| 4 | 0.634±0.577 | 0.773±0.586 | 0.863±0.788 | 1.052±0.830 | 95.2% | 91.0% | 99.8% | 94.6% | 99.1% | 83.0% |
| 5 | 0.736±0.481 | 1.072±1.229 | 0.964±0.675 | 1.350±1.493 | 96.0% | 93.1% | 99.9% | 94.2% | 99.9% | 76.0% |
| 6 | 0.529±0.436 | 0.498±0.272 | 0.771±0.668 | 0.686±0.459 | 95.7% | 96.2% | 100.0% | 92.9% | 99.6% | 56.9% |
| 7 | 0.786±0.514 | 0.789±0.639 | 1.047±0.706 | 1.068±0.888 | 96.7% | 93.4% | 99.7% | 92.0% | 98.5% | 67.7% |
| 8 | 0.734±0.396 | 0.766±0.476 | 1.002±0.559 | 1.010±0.629 | 95.2% | 94.1% | 99.7% | 96.0% | 98.1% | 58.9% |
| 9 | 0.452±0.256 | 0.532±0.478 | 0.629±0.381 | 0.711±0.566 | 94.6% | 92.8% | 100.0% | 96.1% | 100.0% | 82.1% |
| 10 | 0.565±0.272 | 0.671±0.565 | 0.745±0.368 | 0.887±0.707 | 96.8% | 93.9% | 99.9% | 96.4% | 99.9% | 70.8% |
| All | 0.633±0.436 | 0.700±0.611 | 0.848±0.600 | 0.929±0.792 | 96.2% | 94.5% | 99.9% | 95.3% | 99.4% | 76.2% |

Moreover, Mann-Whitney U-tests demonstrated statistically significant differences between ETLD and OTLD across all metrics (MAE: $p < 0.001$; RMSE: $p < 0.001$; CC: $p < 0.001$), thereby validating the efficacy of our enhancements to the TLD framework.

We conducted an in-depth assessment of tracking precision by employing a one-pass evaluation across varying location error thresholds, a method commonly referenced to as the precision plot in object tracking(47). A corresponding recall plot was also generated, as depicted in Figure 4. This analysis reveals that the ETLD reliably identifies a broader range of targets across diverse error thresholds, outperforming the OTLD by successfully tracking targets that the latter fails to detect. A location error threshold of 20 was systematically chosen to determine the definitive precision and recall values, as exemplified in Figure 3 and Table 2.

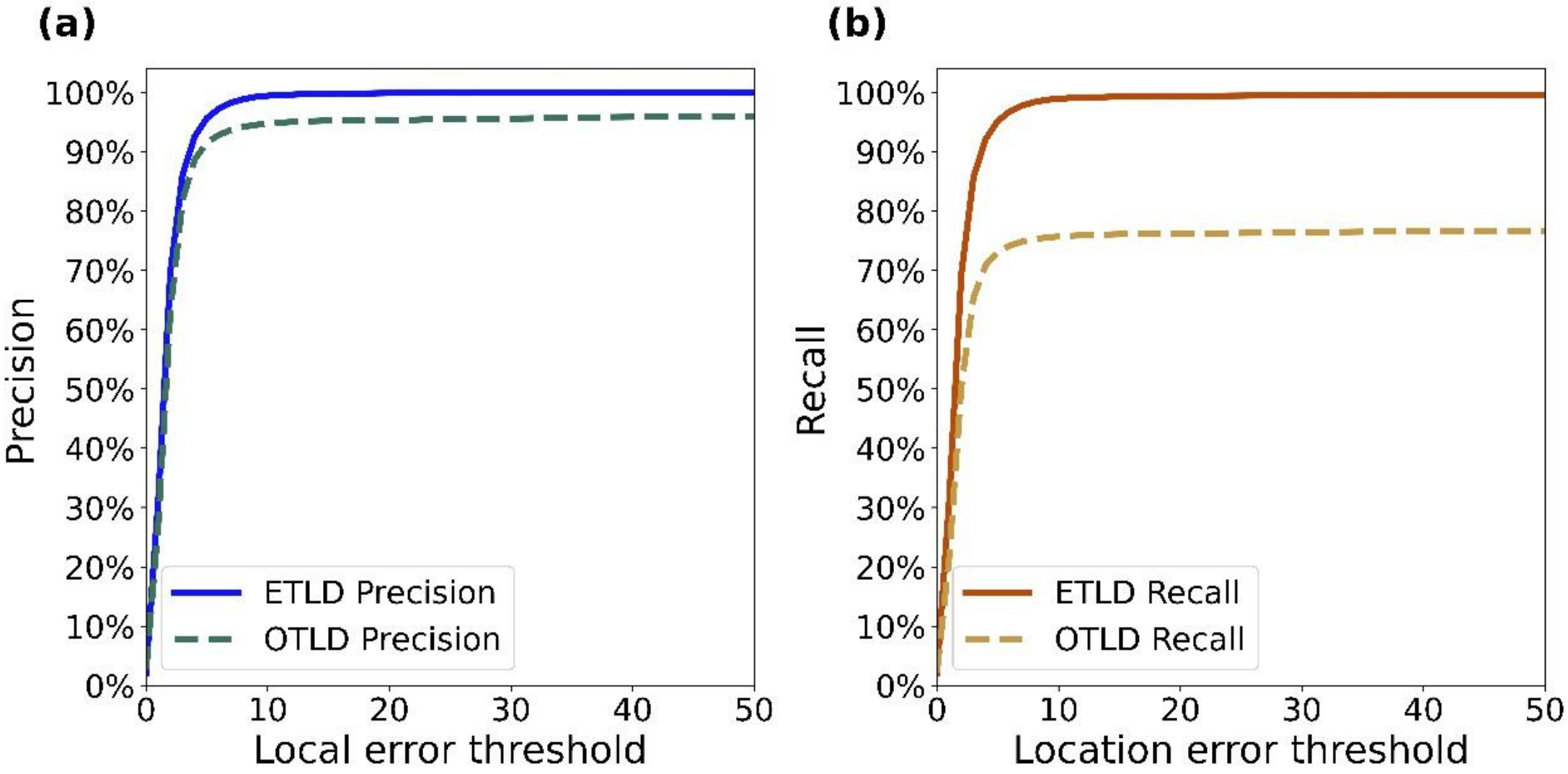

**Figure 4. The precision and recall plot of the ETLD and OTLD methods. (a) The precision plot. (b) The recall plot.**

### 3.2. Accuracy of Segmentation

Given that the segmentation process is contingent upon tracking results, and considering ETLD's significantly superior tracking performance compared to OTLD, we refrained from integrating OTLD with the ICV method. Figure 5 elucidates the segmentation performance achieved by the synergistic application of the ETLD+ICV method. In Figure 5(a), the segmentation results for a sample patient across various treatment fractions are exhibited, with the ground truth demarcated by red outlines and the algorithmic predictions indicated by blue outlines. A close concordance between computed and reference outlines was observed, substantiating the algorithm's spatial fidelity. Figure 5(b) summarizes the statistical aggregation of segmentation outcomes across all enrolled patients. Notably, the Dice global values surpass 82% across the board, with a mean Dice global value of 87.7%. These results attest to the robust performance of the proposed automatic segmentation protocol, affirming its efficacy in delineating the target volume with high precision.

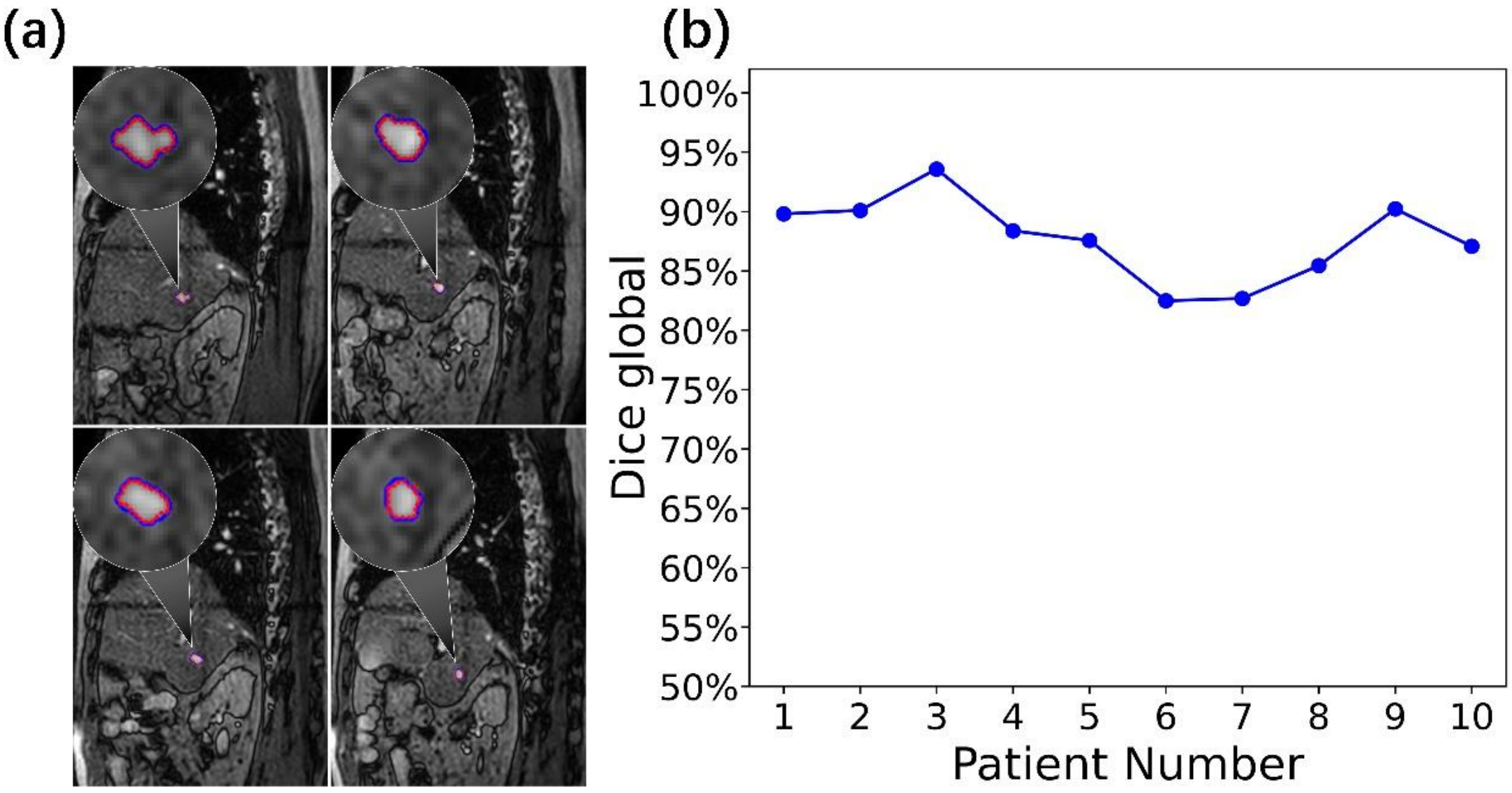

**Figure 5. Segmentation accuracy assessment of the ETLD+ICV method. (a) Direct comparison of segmentation masks for a sample patient, with ground truth in red and predictions in blue. The circular region represents the magnified view, with a magnification factor of 4. (b) Compilation of Dice global values, evidencing the segmentation accuracy across all test patients (n=10).**

### 3.3. Computational Efficiency

Our experiments were conducted in batches, following the hardware specifications detailed in Section 2.5. Even without GPU acceleration, the method maintained computational efficiency. After excluding the initialization phase, the mean processing speeds were 22.304 ± 7.733 FPS for tracking and 25.180 ± 16.116 FPS for segmentation.

It is important to note that variability in processing time was observed across the patient cohort due to inter-patient anatomical heterogeneity. Nevertheless, the overarching average processing time per frame was calculated to be under 50 ms for tracking and below 40 ms for segmentation. These figures demonstrate that our method is highly efficient, with a total processing time of under 100 milliseconds per frame. This level of performance underscores the method's suitability for real-time radiotherapy applications, where swift and seamless processing is paramount.

**4. Discussion**

MRIgRT excels in precision radiotherapy by offering superior soft tissue contrast and eliminating the need for marker implantation(15). To enhance target volume motion tracking during MRIgRT, we developed a novel tracking method termed the ETLD, which builds upon the foundational TLD framework and incorporates three critical enhancements. In addition, ETLD was integrated with the ICV model to enable precise delineation of the target volume.

The OTLD algorithm, known for its high-performance tracking, has been preliminarily used for MRI target tracking(35). However, it encountered challenges with tracking failures and inaccuracies due to varying image quality between natural imaging scenarios and cine-MRI images(35), a common issue in motion tracking for MRIgRT. Moreover, small anatomical structures further exacerbated the difficulty of achieving reliable tracking. We developed the ETLD by enhancing the OTLD algorithm to address the stringent precision requirements in radiotherapy. Enhancements included search process optimization, an enhanced median-flow tracker, and dynamic detection region adjustments. Experiments utilized MRI data from higher magnetic field strengths (1.5T) to mitigate image quality biases and performed the necessary preprocessing. Low-quality data were filtered out through an NRIQA model and manual verification. Additionally, since the OTLD algorithm provides only location predictions, we integrated the ICV segmentation method with the ETLD to ensure precise target volume coverage, termed ETLD+ICV. This innovative approach combines tracking and segmentation, overcoming the limitations of focusing solely on one aspect.

The ETLD's tracking performance was validated using a 3.5D MRI dataset containing 106,000 frames from 77 treatment fractions. Results demonstrated MAE below 0.8 mm, RMSE below 1.05 mm, CC exceeding 94.3%, precision over 99%, and recall above 98% across all patients. In contrast, OTLD exhibited significantly elevated tracking errors and lower precision and recall, with 158 complete tracking failures observed in cine-MRI sequences. Compared to a previous similar study(35) (>1.5mm tracking error, >98% precision, >95% recall), our method achieved enhanced performance on a substantially larger dataset (106,000 frames versus 14,467 frames). These comparative results highlight the effectiveness of the proposed enhancements in improving OTLD's motion tracking capabilities. We attribute these improvements to the optimized sliding window and narrowed scaling factor, which focus the search process on small targets, while the neighborhood size set based on prior information enhances the tracker’s adaptability to rapid target displacements caused by irregular respiratory motion. Finally, the dynamically updated detection regions enable the algorithm to discard potential erroneous information from detection, thereby providing high-quality positional predictions.

Meanwhile, the ETLD+ICV method achieved Dice global scores exceeding 82% across all patients, with a mean of 87.7%, demonstrating precise target volume coverage and validating the high accuracy of ETLD-based motion tracking. The measured average tracking latency remained below 50 ms, and the total system latency, including segmentation processing, was consistently under 100 ms, fully meeting the requirements of the AAPM Task Group 264, which defines "real-time" as a system latency below 500 ms(48).

Real-time MRIgRT is emerging. In early studies on target volume motion tracking, Yun J et al. applied a PCNN for lung tumor tracking, achieving tracking errors below 3.5 mm(23, 24). However, this method requires additional manual operations and employs patient-specific PCNN parameters, leading to an unavoidably lengthy pre-training process. Bourque AE and his team combined a particle filter algorithm with an autoregressive model to achieve millimeter-level tracking accuracy for lung tumors on 1.5T MRI images. However, their approach introduced subjective errors during tracking(25, 26). Seregni M et al. integrated optical flow with template matching to attain a median tracking error of 1 mm in liver motion tracking, yet they also reported compromised algorithm robustness due to static template libraries(27). In contrast, the ETLD receives cine-MRI sequence input and directly outputs predicted positions, eliminating additional subjective operations. Furthermore, its online-updated detector enables respiratory motion adaptation, achieving high tracking precision and recall.

Since most studies have not made their source code publicly available, this study did not conduct direct quantitative comparisons with other methods, instead analyzing the advantages of our method based on the best performance data reported in the literature. Despite differences in experimental conditions, such comparisons remain informative. Compared to recent DL-based motion tracking methods(28-31), the primary advantage of ETLD+ICV lies in its elimination of pre-training requirements, thereby circumventing the need for large-scale, high-quality datasets that remain challenging in medical imaging. This is mainly due to the characteristics of the OTLD framework, as reflected by the online updating of the detector and object model, which enables adaptation to irregular respiratory motion. Simultaneously, the median-flow tracker ensures consistent and precise trajectory predictions. However, algorithm optimization is still essential. The improved detector and tracker have been proven to bring significant performance enhancements, as shown in Table 2. Furthermore, its deployment does not require high-performance GPUs, indicating potential for cost-effective implementation. On the other hand, the proposed method could track the motion of different anatomical structures, such as the entire liver. Although further validation is necessary, the demonstrated sub-millimeter tracking error in our 3.5D MRI dataset with small targets exceeding 97.9% coverage suggests potential applicability across target dimensions, as tracking small targets is typically more challenging than monitoring general-sized targets(49). Additionally, the ETLD can accept continuous 2D image sequences from other imaging modalities as input, extending its applicability beyond MRI-Linac systems to scenarios such as intraoperative radiotherapy and ultrasound-guided biopsies, particularly in proton therapy contexts.

Proton therapy faces several challenges related to daily imaging guidance: (1) visibility of many tumors, (2) real-time tracking of moving tumors during dose delivery, and (3) anatomical variations and organ motion. Combining MRI with proton therapy promises precise and effective dose implementation. Our ETLD+ICV tracking method offers a potentially preferable solution for improving targeting accuracy in proton therapy.

The primary limitation of this study arises from the lack of clinical validation within real-world radiotherapy workflows, constrained by proprietary technical interface limitations of the MRI-Linac system. While our method was evaluated on 1.5T MRI data, its robustness across varying magnetic field strengths, such as 0.35T MRI, remains unverified. Moreover, the absence of standardized benchmarking datasets, compounded by heterogeneous imaging protocols and equipment variations, hinders objective algorithm comparisons and underscores the critical need for multi-institutional collaborative data sharing initiatives. Additionally, although our algorithm reduces tracking errors, the sliding window approach can be resource-intensive, impacting clinical applicability. Future work will focus on these limitations by integrating novel architectures to further enhance tracking accuracy and efficiency, while prioritizing validation of the algorithm's robustness against interference in low-quality data and quantitative assessment of its clinical performance. Moreover, potentially applying the method to other imaging scenarios like infrared scenes will expand its applicability.

## 5. Conclusion

This study successfully developed and validated an automatic real-time markerless motion tracking method using the ETLD+ICV for MRIgRT. The method's high precision and recall, achieved without pre-training, surpass current techniques and offer pronounced improvements for MRIgRT, enhancing radiotherapy effectiveness. The implications for radiotherapy are that advanced tracking systems may improve patient outcomes. Future work will investigate further enhancements to the tracking algorithm and explore its application in other imaging contexts.

## 6. Declarations

### 6.1. Ethics approval and consent to participate

This study was conducted in compliance with the Declaration of Helsinki – Ethical Principles for Medical Research Involving Human Participants. This study approval from the National Cancer Center/Cancer hospital,Chinese Academy of Medical Science and Peking Union Medical College, National GCP Center for Anticancer Drugs, The Independent Ethics Committee was obtained prior to the commencement of the study (NCC2020C-479). All patient data used in this study were obtained with written consent from the patients or their guardians.

### 6.2. Consent for publication

Not applicable.

### 6.3. Availability of data and materials

The datasets generated and/or analysed during the current study are not publicly available due to the data security requirements of Cancer Hospital, Chinese Academy of Medical Sciences, but are available from the corresponding author on reasonable request.

### 6.4. Competing interests

The authors declare no competing interests.

### 6.5. Funding

This work was supported by the National Natural Science Foundation of China (No. 12105368 and 62475020), the Cooperation Fund of CHCAMS and the National High-Level Hospital Clinical Research Funding (No. CFA202502022).

### 6.6. Authors' contributions

JY-C: Conceptualization, Data curation, Formal analysis, Investigation, Resources, Project

administration, Software, Visualization, Writing -review & editing, Funding acquisition, Supervision. GH-W: Project administration, Formal analysis, Supervision, Writing -review & editing. SQ-C: Data curation, Formal analysis, Investigation, Software, Visualization, Writing - original draft, Writing -review & editing. ZL-W: Data curation, Investigation, Software, Visualization, Writing - original draft. JR-D: Conceptualization, Resources, Supervision Writing -review. Y-T: Data curation, Resources, Supervision Writing -review. SR-Q: Data curation, Writing -review Y-C: Data curation, Writing -review Resources, RA-Z: Data curation, Writing -review.

### 6.7. Acknowledgements

Not applicable.